\documentstyle[aps,preprint,amssymb,epsf,tighten]{revtex}
\begin{document}
\draft

\title{High-powered Gravitational News}

\author{Nigel T. Bishop${}^{1}$,
        Roberto G\'omez${}^{2}$,
	Luis Lehner${}^{2}$, \\
        Manoj Maharaj${}^{3}$ and
        Jeffrey Winicour${}^{2}$}
\address{
${}^{1}$Department of Mathematics, Applied Mathematics and Astronomy,\\
University of South Africa, P.O. Box 392, Pretoria 0003, South Africa \\
${}^{2}$Department of Physics and Astronomy,\\
University of Pittsburgh, Pittsburgh, PA 15260\\
${}^{3}$Department of Mathematics and Applied Mathematics,\\
University of Durban-Westville, Durban 4000, South Africa }

\maketitle

\begin{abstract}

We describe the computation of the Bondi news for gravitational
radiation.  We have implemented a computer code for this problem. We
discuss the theory behind it as well as the results of validation
tests. Our approach uses the compactified null cone formalism, with the
computational domain extending to future null infinity and with a
worldtube as inner boundary. We calculate the appropriate full Einstein
equations in computational eth form in (a) the interior of the
computational domain and (b) on the inner boundary. At future null
infinity, we transform the computed data into standard Bondi
coordinates and so are able to express the news in terms of its
standard $N_{+}$ and $N_{\times}$ polarization components. The
resulting code is stable and second-order convergent. It runs
successfully even in the highly nonlinear case, and has been tested
with the news as high as 400, which represents a gravitational
radiation power of about $10^{13}M_{\odot}/sec$.

\end{abstract}

\section{Introduction}
\label{intro}

In a series of papers in the 1960s Bondi and
others~\cite{bondi,sachs,Penrose,NP,tam} developed the theory of
gravitational radiation at null infinity, and this formalism has now
become accepted practice: theoretical results about gravitational
radiation at future null infinity (${\cal I}^+$) are usually expressed
in terms of the Bondi news function $N$. However, until now it has not
been possible to calculate the news numerically, except in spacetimes
with special symmetries.  This paper describes the theory, as well as a
code, for computing the news in general (asymptotically flat)
spacetimes and in regimes that include the highly nonlinear case. The
code runs with the news $N$ as large as $400$ (in geometric units in
which $N$ is dimensionless). This is enormous and means that the code
can cope with a power output of order $N^2 c^5/G \approx 10^{60}W$ in
conventional units.  This exceeds the power that would be produced if,
in 1 second, the whole Galaxy were to become gravitational radiation.

Most work in numerical relativity is done in the Cauchy ``3 + 1''
formalism, with the gravitational radiation estimated by perturbative
methods~\cite{ab1,ab2,ab3}; these methods have not been tested on
high-powered waveforms.  However, here we are using the characteristic,
or null cone, formalism with radial distance compactified so that
${\cal I}^+$ is contained in the (finite) grid. The theoretical
foundations for this approach were laid in the
1980s~\cite{Stewart,isaac,newt,nullinf}. Numerically, it has been
implemented for the model problem of the scalar wave
equation~\cite{Gom} and in general relativity under various restrictive
conditions~\cite{Bis90,Clarke,papa,cce}.

In a previous paper~\cite{cce} we were able to compute the news in the
quasi-spherical case, i.e. with the nonlinear terms in Einstein's
equations ignored. The notation and approach here follow that
in~\cite{cce}. The computational domain is bounded by an inner
(timelike or null) worldtube, and the outer boundary of the domain is
${\cal I}^+$. The computational domain is foliated into a sequence of
null cones with initial data given on the first cone. Appropriate data
on the inner boundary then enable us to use Einstein's equations to
find the spacetime geometry throughout the computational domain (see
figure 1 in~\cite{cce}). In addition to its other contributions, this
work is also an important step towards the full implementation of
Cauchy-characteristic matching as envisaged in\cite{cce} (see
also~\cite{Bis92,Bis93,Ccpaper,Bis95,dinvic}), since the characteristic
code is now {\em complete}.

The first step in this paper is the calculation of the nonlinear terms
in the Einstein equations for the Bondi-Sachs metric, and the
expression of these terms in computational eth form~\cite{eth,penrin}.
This is done computationally using a symbolic programming language, and the results are
described in Section~\ref{ncf}. The inner boundary of the computational
domain may be a timelike, or an incoming null, worldtube, and
Section \ref{sec:cons} discusses the required boundary data~\cite{tam}.
Next, in Section~\ref{asymp}, we investigate behavior at the outer
boundary ${\cal I}^+$. We use null cone coordinates that are defined
quite generally in terms of a $2+1$ decomposition of the inner world
tube. Consequently, they do not form an inertial Bondi system at ${\cal
I}^+$. However, by carrying out a transformation to an inertial frame,
we express the news in terms of the two standard polarization modes
$N_{+}$and $N_{\times}$\cite{mtw}. Using computer algebra, these
expressions are translated into computational eth form.

At this stage of the paper, the theoretical background for the problem
has been completely specified. Section~\ref{sec:num} considers various
details of the numerical implementation, in particular those aspects
that did not occur in the quasispherical case. We go on in
Section~\ref{sec:test} to describe the results of testing the code for
stability and accuracy  and demonstrate that the code is second-order
convergent.

We also perform a series of runs for an incoming asymmetric
gravitational pulse incident on a Schwarzschild black hole. This is the
fully nonlinear version of the classic black hole scattering problem
first worked out in the perturbative regime~\cite{price}. As the
magnitude of the pulse is ramped up various nonlinear effects become
evident. The largest pulse investigated leads to a peak news of $N=400$
for the gravitational radiation backscattered to ${\cal I}^+$.

Some of the computational eth expressions found in this paper are quite
long, and they are given in Appendices: Appendix A for the nonlinear terms
in the Einstein equations, and Appendix B for the news.

\section{The null cone formalism}
\label{ncf}

First we summarize our notation and previous results about the Einstein
equations in the null cone formalism. These earlier results were
complete only in the quasispherical approximation~\cite{cce}. We next
proceed to discuss the computer algebra techniques that were used to
extend these results to the full nonlinear case in computational eth
form~\cite{eth}, and we present the completed results.

\subsection{Previous results}
\label{ncold}

We use coordinates based upon a family of outgoing null hypersurfaces.
We let $u$ label these hypersurfaces, $x^A$ $(A=2,3)$, label
the null rays and $r$ be a surface area coordinate. In the resulting
$x^\alpha=(u,r,x^A)$ coordinates, the metric takes the Bondi-Sachs
form~\cite{bondi,sachs}
\begin{eqnarray}
   ds^2 & = & -\left(e^{2\beta}(1 + {W \over r}) -r^2h_{AB}U^AU^B\right)du^2
        -2e^{2\beta}dudr -2r^2 h_{AB}U^Bdudx^A \nonumber \\
        & + & r^2h_{AB}dx^Adx^B,    \label{eq:bmet}
\end{eqnarray}
where $W$ is related to the more usual Bondi-Sachs variable $V$ by
$V=r+W$; and where $h^{AB}h_{BC}=\delta^A_C$ and
$det(h_{AB})=det(q_{AB})$, with $q_{AB}$ a unit sphere metric.  In
analyzing the Einstein equations, we also use the intermediate variable
\begin{equation}
Q_A = r^2 e^{-2\,\beta} h_{AB} U^B_{,r}.
\end{equation}

We work in stereographic coordinates $x^A=(q,p)$ for which the unit sphere
metric is
\begin{equation}
q_{AB} dx^A dx^B = \frac{4}{P^2}(dq^2+dp^2),
\end{equation}
where
\begin{equation}
        P=1+q^2+p^2.
\end{equation}
We express $q_{AB}$ in terms of a complex dyad $q_A$ (satisfying $q^Aq_A=0$,
$q^A\bar q_A=2$, $q^A=q^{AB}q_B$, with $q^{AB}q_{BC}=\delta^A_C$ and
$ q_{AB} =\frac{1}{2}\left(q_A \bar q_B+\bar q_Aq_B\right)$). We fix
the dyad by the explicit choice
\begin{equation}
      q^A=\frac{P}{2}(1,i)
\end{equation}
with $i=\sqrt{-1}$.  Note that we have departed from other
conventions~\cite{penrin} to avoid factors of $\sqrt{2}$ which are
awkward in numerical work.  For an arbitrary Bondi-Sachs metric,
$h_{AB}$ can then be represented by its dyad component $J=h_{AB}q^Aq^B
/2$, with the spherically symmetric case characterized by $J=0$. The
full nonlinear $h_{AB}$ is uniquely determined by $J$, since the
determinant condition implies that the remaining dyad component
$K=h_{AB}q^A \bar q^B /2$ satisfies $1=K^2-J\bar J$.  We also introduce
spin-weighted fields $U=U^Aq_A$ and $Q=Q_Aq^A$, as well as the (complex
differential) eth operators $\eth$ and $\bar \eth$. Tensor quantities
and spin-weighted quantities are related as follows:
\begin{eqnarray}
h_{22}=\frac{2}{P^2}(J+\bar J+2K),\;h_{23}=h_{32}=\frac{2}{P^2}(\bar J-J)i,\;
h_{33}=\frac{2}{P^2}(2K -J-\bar J), \nonumber \\
U^2=\frac{P}{4}(U+\bar U),\;U^3=\frac{P}{4}(\bar U-U)i,\;
Q_2=\frac{Q+\bar Q}{P},\;Q_3=\frac{i}{P}(\bar Q -Q).
\label{eq:huq}
\end{eqnarray}
Refer to {}~\cite{eth,cce} for further details.

The Einstein equations $G_{\mu\nu}=0$ decompose into hypersurface
equations, evolution equations and conservation laws. In writing the
field equations, we follow the formalism given in~\cite{newt,nullinf}.
We find for the hypersurface equations:
\begin{eqnarray}
      \beta_{,r} &=& N_\beta, 
   \label{eq:beta} \\
          U_{,r}  &=& r^{-2}e^{2\beta}Q +N_U, 
     \label{eq:wua} \\
     (r^2 Q)_{,r}  &=& -r^2 (\bar \eth J + \eth K)_{,r}
                +2r^4\eth \left(r^{-2}\beta\right)_{,r} + N_Q, 
     \label{eq:wq} \\
W_{,r} &=& \frac{1}{2} e^{2\beta}{\cal R} -1
- e^{\beta} \eth \bar \eth e^{\beta}
+ \frac{1}{4} r^{-2} \left(r^4
                           \left(\eth \bar U +\bar \eth U \right)
                     \right)_{,r} + N_W,
 \label{eq:ww}
\end{eqnarray}
where~\cite{eth}
\begin{equation}
{\cal R} =2 K - \eth \bar \eth K + \frac{1}{2}(\bar \eth^2 J + \eth^2 \bar J)
          +\frac{1}{4K}(\bar \eth \bar J \eth J - \bar \eth J \eth \bar J).
     \label{eq:calR}
\end{equation}
Next, the evolution equation takes the form
\begin{eqnarray}
    && 2 \left(rJ\right)_{,ur}
    - \left(r^{-1}V\left(rJ\right)_{,r}\right)_{,r} = \nonumber \\
    && -r^{-1} \left(r^2\eth U\right)_{,r}
    + 2 r^{-1} e^{\beta} \eth^2 e^{\beta}- \left(r^{-1} W \right)_{,r} J
    + N_J.
    \label{eq:wev}
\end{eqnarray}
The remaining independent equations are the conservation conditions, which
are discussed in Section \ref{sec:cons}.

In the above equations $N_\beta$, $N_U$, $N_Q$, $N_W$ and $N_J$
represent the nonlinear aspherical terms (in a sense specified in ~\cite{cce}).
Expressions for these terms are known~\cite{cce} in the form of tensors
and covariant derivatives, rather than complex spin-weighted fields and
$\eth$. We now calculate those nonlinear terms in the desired form.

\subsection{New results}
\label{ncnew}

We have used computer algebra to calculate the nonlinear terms in two
separate ways. In the first approach we start from the expressions
reported previously ~\cite{cce}. In the second approach we start from
the text-book definition of the Ricci tensor
\begin{equation}
R_{\mu \nu}=\partial_\alpha \Gamma^\alpha_{\mu \nu}
-\partial_\nu \Gamma^\alpha_{\mu\alpha}
+\Gamma^\alpha_{\sigma\alpha} \Gamma^\sigma_{\mu\nu}
-\Gamma^\alpha_{\sigma\nu} \Gamma^\sigma_{\mu\alpha}.
\end{equation}
We have checked that the two approaches gave the same results by
substituting the total expressions (including the quasispherical and
nonlinear parts) obtained for $\beta_{,r}$, $U_{,r}$, $Q_{,r}$,
$W_{,r}$ and $J_{,ur}$ into the relevant part of the Ricci tensor, and
confirming that the difference between the two results was zero.

The translation of a formula from tensor to eth formalism is ideal for
computer algebra: it is straightforward and algorithmic, but lengthy.
The first step is to expand any covariant derivatives in terms of
partial derivatives and connection terms. Then using
equation (\ref{eq:huq}), each $h_{AB}$ is expressed in terms of $J$,
$\bar J$ and $K$, and the $U^A$ and $Q_A$ are expressed in terms of $U$
and $\bar U$, and $Q$ and $\bar Q$, respectively.  At this stage
everything is in terms of spin-weighted fields, but angular derivatives
of these fields are partial derivatives with respect to the
coordinates. The next step is to transform these partial derivatives to
$\eth$ and $\bar \eth$ operators. We use simple linear algebra applied
to earlier results~\cite{eth} to obtain
\begin{equation}
A_{,q}=\frac{\eth A +\bar \eth A - 2ipsA}{P},\;
A_{,p}=\frac{i(\bar \eth A -\eth A + 2qsA)}{P}
\end{equation}
where $A$ is a spin-weighted field with spin-weight $s$. For convenient
reference we list all the spin-weighted fields that are used in the
Bondi-Sachs metric and their spin-weights:
\begin{tabbing}
Spin-weight: x\= -2xxxxxxxxxx\= -1xxxxxxxxxx\= 0xxxxxxxxxx\= 1xxxxxxxxxx\= 2 \kill
Spin-weight: \> -2 \> \,\,\,-1 \> \,\,\,\,\,\,\,\,\,\,0 \> \,\,\,\,\,1 \> 2 \\
Field: \> $\bar J$ \> $\bar Q$, $\bar U$ \> $\beta$, $K$, $W$ \> $Q$, $U$
\> $J$
\end{tabbing}
Note further that the operator $\eth$ increases the spin-weight by 1 and
the operator $\bar \eth$ decreases the spin-weight by 1. For example,
$\eth \beta$, $\bar \eth J$ and $\eth^2 \bar U$ all have spin-weight 1.

For quantities that are not spin-weight 0, the expressions for $A_{,q}$
and $A_{,p}$ involve $q$ and $p$. However, the eth representation of
geometric spin-weighted quantities (e.g.  $R_{rA}q^A$) should not
involve $q$ and $p$, so if they appear due to one partial derivative
they should be cancelled out by something else in the expression; also
$P$ should not appear in the final representation. These features
provide useful checks in computer algebra calculations.

The procedures described above are complete for quantities that involve
{\em one} angular derivative, including second derivatives involving
one angular derivative and one derivative with respect to $u$ or $r$.
Quantities involving {\em two} angular derivatives are unpacked one
derivative at a time, as illustrated in the  example
\begin{eqnarray}
W_{,qp}=(W_{,q})_{,p}=\bigg( \frac{\eth W + \bar \eth W}{P} \bigg)_{,p}
=\frac{P((\eth W)_{,p} + (\bar \eth W)_{,p})-2p(\eth W + \bar \eth W)}{P^2}
\nonumber \\
=\frac{i(\bar \eth \eth W -\eth^2 W +2 q \eth W
+ \bar \eth^2 W -\eth \bar \eth W -2 q \bar \eth W)
-2p(\eth W + \bar \eth W)}{P^2}
\end{eqnarray}

We present the results of our computer algebra calculations in Appendix
\ref{app:appa}. Then the complete set of null cone hypersurface and
evolution equations is given by equations (\ref{eq:beta}) to
(\ref{eq:wev}), augmented by equations (\ref{eq:nbeta}) to
(\ref{eq:nji}). These equations are suitable for numerical
implementation in computational eth form.

Most of the computer algebra was done using Maple, with some developmental
work and checking on Mathematica and REDUCE. The Maple scripts
are available on the World Wide Web (http://artemis.phyast.pitt.edu/hpgnews)

\section{Conservation conditions}
\label{sec:cons}

Our discussion up to this point has only addressed the six hypersurface
and evolution equations, designated by Bondi as the ``main'' Einstein
equations~\cite{bondi}. They correspond to the six components of the
Ricci tensor itemized in Appendix \ref{app:appa}. Together with the
equations $R^r_{\alpha}=0$, they form a complete set of components of
the vacuum Einstein equations. Given that the main equations are
satisfied, Bondi used the Bianchi identities to show that one of these,
the ``trivial'' equation $R^r_r=0$, is automatically satisfied. He
further showed that the remaining three equations
\begin{equation}
          R^r_u=0,\;q^AR^r_A=0.
\label{eq:cons}
\end{equation}
are satisfied on a complete outgoing null
cone if they are satisfied on a single spherical cross-section.  By
choosing this sphere to be at infinity, he identified these three
equations as conservation conditions for energy and angular momentum.

In the context of our present work, we obtain a solution by requiring
that these conservation conditions be satisfied on a worldtube $\Gamma$
forming the inner boundary of the characteristic domain~\cite{tam}.
These conditions are automatically satisfied when matching across
$\Gamma$ to an interior solution generated, say, by Cauchy evolution.
Matching is anticipated to be a primary application of the
characteristic algorithm. However, in carrying out a purely
characteristic evolution of the exterior, based upon data on $\Gamma$
and on an initial null hypersurface, these conservation conditions must
be imposed separately. This applies to the stand-alone tests of the
characteristic code presented in Section \ref{sec:test}. An alternative
is to take the limit in which the worldtube collapses to a world line,
in which case the conservation conditions reduce to regularity
conditions at the vertices of the null cones. This approach was used in
numerical evolution with an axisymmetric characteristic code~\cite{papa}
but would be computationally prohibitive in the 3-dimensional case.

The conservation conditions (\ref{eq:cons}) dictate which metric
variables can be given freely on $\Gamma$ and which are subject to
constraints. They contain no second $r$-derivatives of the metric and
are the analogues, with respect to an $r$-foliation, of the traditional
momentum constraints in a Cauchy formalism. We can therefore express
them as
\begin{equation}
     D^j (K_{ij}-\gamma_{ij} K)=0 \label{eq:momc}
\end{equation}
in terms of the intrinsic metric $\gamma_{ij}$, the associated
covariant derivative $D_i$
and the extrinsic curvature $K_{ij}$ in the induced coordinates
$x^i=(u,x^A)$ of the worldtube $\Gamma$.

From (\ref{eq:bmet}), The intrinsic metric is
\begin{equation}
   \gamma_{ij}dx^i dx^j =-e^{2\beta}{V \over r}du^2
        +r^2h_{AB}(dx^A-U^Adu)(dx^B-U^Bdu).
\end{equation}
In analogue to the $3+1$ decomposition of the Cauchy formalism, a
$2+1$ decomposition of the timelike worldtube geometry leads to the identification of
$g_{AB}=r^2h_{AB}$ as the metric of the 2-surfaces of constant $u$ which
foliate the world-tube, $e^{2\beta}V/r$ as the square of the lapse
function and $(-U^A)$ as the shift vector. The extrinsic
curvature of $\Gamma$ is
\begin{equation}
    K_{ij} = \delta_i^{\alpha} \delta_j^{\beta}
                \nabla_{\alpha} n_{\beta}
\end{equation}
in terms of the unit normal
$n_{\alpha}=(r/V)^{\frac{1}{2}}e^{\beta}\nabla_{\alpha} r$.

Part of the worldtube data is specification of the gauge. One simple
choice is to set the shift to zero ($U^A=0$), and the lapse to one
($V/r=e^{-2\beta}$). In this case, the components of the extrinsic
curvature are
\begin{equation}
    K_{uu}=2\beta_{,u} +{1\over 2}e^{-2\beta}g_{uu,r},
\end{equation}
\begin{equation}
    K_{uA}=\beta_{,A} +{1\over 2}e^{-2\beta}g_{uA,r},
\end{equation}
\begin{equation}
    K_{AB}={1\over 2} (-g_{AB,u}+e^{-2\beta}g_{AB,r}),
\end{equation}
with
\begin{equation}
    K=-2\beta_{,u}+{2\over r} e^{-2\beta}
         -{1\over 2} e^{-2\beta}g_{uu,r},
\end{equation}
where $g_{uu}$, $g_{uA}$ and $g_{AB}$ are obtained from (\ref{eq:bmet}).
It is then easy to work out the specific form of the conservation
conditions using the identity
\begin{equation}
        D^j K_{ij} ={1\over \sqrt{-\gamma}}
                        \partial_j(\sqrt{-\gamma} K_i^j)
                  +{1\over 2}K_{mn}{\gamma^{mn}}_{,i}  \, \, ,
\end{equation}
which holds for any symmetric tensor.

In order to elucidate the content of these equations we lump under the
symbol ${\cal K}_{\beta}$ or ${\cal K}_A$ any  purely null-hypersurface
term (composed out of $\beta$, $U^A$, $V$ and $h_{AB}$ and their $r$
and $x^A$ derivatives) together with any purely worldtube term (composed
out of $h_{AB}$ and its $u$ and $x^A$ derivatives).  Then the
$u$-component of (\ref{eq:momc}) takes the form
\begin{equation}
    \beta_{,u}={\cal K}_{\beta};
\label{eq:betacon}
\end{equation}
and the $x^A$-components are
\begin{equation}
    (e^{-2\beta} g_{uA,r} -2\beta_{,A})_{,u}={\cal K}_A,
\end{equation}
which may be re-expressed in computational eth form as
\begin{equation}
Q_{,u}=-2\eth \beta_{,u}-q^A{\cal K}_A.
\label{eq:Qcon}
\end{equation}

Equations (\ref{eq:betacon}) and (\ref{eq:Qcon}) are generalizations to
a finite worldtube of the energy and angular momentum conservation laws
found by Bondi. They imply that $\beta$ and $Q$ cannot be given freely
on the the worldtube but are subject to constraints. Since the
worldtube values of $\beta$ and $Q$ enter as integration constants in
the hypersurface equations, the evolution scheme must either update
their values in accord with equations (\ref{eq:betacon}) and
(\ref{eq:Qcon}) or prescribe them by matching to an interior solution.
(In the test cases presented in Section \ref{sec:test} we match to an
analytic interior solution).  The remaining integration constants can
then be posed freely on the worldtube. This gives the formal result (in
our present choice of gauge) that specification of $h_{AB}$ on the
initial null hypersurface and on the worldtube and specification
of $\beta$ and $Q$ on the initial slice of the worldtube determine the
future evolution of the system in the region outside the worldtube.

A similar nullcone-worldtube evolution scheme holds for a general
choice of lapse and shift, although the detailed equations are
considerably more complicated. An example where the worldtube
is null is presented in Section \ref{sec:scat}.

\section{Null infinity} \label{asymp}

The spacetime can be conformally compactified in terms of the metric
$d{\hat s}^2=\ell^2 ds^2$, where $\ell=1/r$.
In $(u,\ell,x^A)$ coordinates, it takes the form
\begin{equation}
   d{\hat s}^2 = -\left(e^{2\beta}V \ell^3 -h_{AB}U^AU^B\right)du^2
        +2e^{2\beta}dud\ell -2 h_{AB}U^Bdudx^A + h_{AB}dx^Adx^B.
   \label{eq:lmet}
\end{equation}
Here $\ell$ is a conformal factor with future null infinity ${\cal
I}^+$ given by $\ell=0$. The physical quantities governing the total
energy and radiation power from the system are constructed from the
leading coefficients (functions of $u$ and  $x^A$) in an expansion of
the metric in powers of $\ell$, in accord with asymptotic flatness at
${\cal I}^+$. The coefficients $H$, $H_{AB}$, $c_{AB}$ and $L^A$ which
enter the construction of the news function are defined by the
expansions, $\beta= H +O(\ell^2)$, $h_{AB}= H_{AB}+\ell
c_{AB}+O(\ell^2)$, $U^A= L^A+2\ell e^{2H} H^{AB}D_B H+O(\ell^2)$ and
$\ell^2 V= D_A L^A+\ell e^{2H}({\cal R}/2 +2D_AD^A H+4D_AH D^A
H)+O(\ell^2)$. (Note here that we have related some of the expansion
coefficients by using the asymptotic content of the Einstein
equations).
 
\subsection{The News}

The gauge freedom in the conformal metric $d {\hat s}^2$ is fixed by
the gauge conditions adopted on the inner worldtube $\Gamma$.  In order
to construct the news function, it is convenient to refer to a
conformal Bondi frame~\cite{tam} with metric $d{\tilde s}^2=\Omega^2
ds^2 =\omega^2 d{\hat s}^2$, with $\Omega=\omega\ell$, which satisfies
the gauge requirements that $Q_{AB}:={\tilde g}_{AB}|_{{\cal
I}^+}=\omega^2 H_{AB}$ is intrinsically a unit sphere metric at ${\cal
I^+}$ and that ${\tilde g}^{\alpha\beta}\nabla_{\alpha} \Omega
\nabla_{\beta} \Omega=O(\Omega^2)$. The conformal Bondi frame
corresponds to an asymptotically Minkowskian inertial frame.
Reference~\cite{quad} discuses the calculation of the news in an
arbitrary conformal frame relative to that in a Bondi frame.

${\cal I^+}$ is geometrically a null hypersurface with null vector
tangent to its generators given by
${\tilde n}^{\alpha}=
{\tilde g}^{\alpha \beta}\nabla_{\beta} \Omega |_{\cal I^+}$
or, equivalently, 
${\hat n}^{\alpha} = {\hat g}^{\alpha\beta}\nabla_{\beta} \ell |_{\cal I^+}$, with
${\tilde n}^{\alpha} =\omega^{-1}{\hat n}^{\alpha}$.
In order to complete a basis for tangent vectors to ${\cal I^+}$,
let $Q^{\alpha}$ be a complex field in the neighborhood of ${\cal I^+}$
satisfying $Q^{\alpha} \nabla_{\alpha}\Omega=O(\Omega)$, ${\tilde
g}_{\alpha\beta}Q^{\alpha}Q^{\beta}=O(\Omega)$ and ${\tilde
g}_{\alpha\beta}Q^{\alpha}{\bar Q}^{\beta}=2+O(\Omega)$.  In the
conformal Bondi frame, the news function can then be expressed in the
form~\cite{high}
\begin{equation}
  N={1\over 2\Omega}Q^{\alpha} Q^{\beta}
     {\tilde \nabla}_{\alpha} {\tilde \nabla}_{\beta}  \Omega 
\label{eq:snews}
\end{equation}
evaluated in the limit of ${\cal I^+}$. (Our conventions are chosen so
that the news reduces to Bondi's original expression in
the axisymmetric case~\cite{bondi}). In terms of
the ${\hat g}_{\alpha\beta}$ frame, with conformal factor ${\hat
\Omega}=\Omega/\omega =\ell$, we then have
\begin{equation}
 N={1\over 2}Q^{\alpha} Q^{\beta}
  ({{\hat \nabla}_{\alpha} {\hat \nabla}_{\beta}{\hat \Omega} \over {\hat \Omega}}
   -\omega{\hat \nabla}_{\alpha} {\hat \nabla}_{\beta} {1\over \omega}).
     \label{eq:lnews}
\end{equation}

We can determine $\omega$ on ${\cal I}^+$ in the ${\hat
g}_{\alpha\beta}$ frame by solving the elliptic equation
governing the conformal transformation of the curvature scalar
intrinsic to the 2-geometry of a $u=constant$ slice of ${\cal I}^+$,
\begin{equation}
     {\cal R}=2(\omega^2+H^{AB}D_A D_B \log \omega),
\label{eq:conf}
\end{equation}
where ${\cal R}$ is defined in equation (\ref{eq:calR}).
The condition that  ${\tilde g}^{\alpha\beta}\nabla_{\alpha} \Omega
\nabla_{\beta} \Omega=O(\Omega^2)$ is consistent with setting
$\partial_{\ell} \omega=0$ but gives a relation for the time dependence
of $\omega$.  Noting that ${\hat g}^{\alpha\beta}\nabla_a \ell
\nabla_{\beta} \ell =e^{-2H}D_AL^A\ell +O(\ell^2)$ and that ${\hat
g}^{\alpha\beta}\nabla_{\beta} \ell={\hat n}^{\alpha}+O(\ell)$, we
obtain $2{\hat n}^{\alpha} \nabla_{\alpha} \log \omega
=-e^{-2H}D_AL^A$. This is used to evolve $\omega$ given a solution of
(\ref{eq:conf}) as initial condition.

In order to obtain an explicit expression for the news (\ref{eq:lnews})
in the ${\hat g}_{\alpha\beta}$ frame we need a representation of
$Q^{\beta}$. The choice is independent of the freedom $Q^{\beta}
\rightarrow Q^{\beta} + \gamma {\tilde n}^{\beta}$, which leaves
(\ref{eq:lnews}) invariant. However, it is important for physical
interpretation to choose the spin rotation freedom $Q^{\beta}
\rightarrow e^{-i\alpha} Q^{\beta}$ to satisfy ${\tilde
n}^{\alpha}{\tilde \nabla}_{\alpha} Q^{\beta}=O(\Omega)$, so that the
polarization frame is parallely propagated along the generators of
${\cal I}^+$. This fixes the polarization modes determined by the real
and imaginary parts of the news to correspond to those of inertial
observers at ${\cal I}^+$. 

We accomplish this by introducing the dyad
decomposition $h^{AB}=(m^A \bar m^B+\bar m^A m^B)/2$ and fixing the
spin rotation freedom of $m^A$ by setting
\begin{equation}
 m^A=\left({\bar J \over J}\right)^{1/4} \Bigg( q^A  \sqrt{ \frac{(K+1)}{2 }  }
    -\bar q^A  J \sqrt{ 1 \over 2(K+1)} \,  \Bigg) ,
\label{eq:mA}
\end{equation}
so that $H^{AB}=(M^A{\bar M}^B+{\bar M}^AM^B)/2$ where $M^A =m^A
+O(\Omega)$. Here, in order to make $m^A$ independent of the phase of
$q^A$ (the basis vector used in the $\eth$ computations), we have
introduced the phase factor $e^{i \gamma}=( \bar J / J)^{1/4}$.  Now,
setting $Q^{\beta}=e^{-i\alpha}\omega^{-1}(0,0,M^A)+\gamma {\tilde
n}^{\beta}$, the requirement of an inertial polarization frame, 
${\tilde n}^{\alpha}{\tilde \nabla}_{\alpha} Q^{\beta}=O(\Omega)$, 
determines the time dependence of the phase $\alpha$. We obtain
\begin{equation}
    2(\partial_u +L^A\partial_A)(i\alpha+\log\omega) =
    +H_{AC}{\bar M}^C (\partial_u M^A
             +L^B\partial_B M^A - M^B\partial_B L^A)
\end{equation}
or, eliminating the time derivative of $\omega$,
\begin{equation}
    2i(\partial_u +L^A\partial_A)\alpha = D_A L^A
    +H_{AC}{\bar M}^C ((\partial_u+L^B\partial_B )M^A
              - M^B\partial_B L^A) .
\label{eq:alphadot}
\end{equation}

Equation (\ref{eq:mA} ) introduces numerical problems in the calculation of
$M^A$ at points where $J=0$. We avoid these by setting
$ M^A = e^{i \gamma}F^A$, where
\begin{equation}
   F^A  = q^A  \sqrt{ \frac{(K+1)}{2 }  }
          -\bar q^A  J \sqrt{ 1 \over 2(K+1)} ,
\end{equation}
Substituting $F^A$ for $M^A$ and setting $\delta=\alpha -\gamma$,
equation (\ref{eq:alphadot}) takes the regularized form
\begin{equation}
    2i(\partial_u +L^A\partial_A)\delta = D_A L^A
     +H_{AC} \bar F^C ( (\partial_u +L^B \partial_B) F^A
             - F^B \partial_B L^A) .
    \label{eq:evphase}
\end{equation}

We can now express the ``inertial'' news (\ref{eq:lnews}) in the ${\hat
g}_{\alpha\beta}$ frame as
\begin{equation}
  N={1\over 2}e^{-2i\alpha}\omega^{-2}M^{\alpha} M^{\beta} 
({{\hat \nabla}_{\alpha} {\hat \nabla}_{\beta} {\hat \Omega} 
    \over {\hat \Omega}}
   -\omega{\hat \nabla}_{\alpha} {\hat \nabla}_{\beta} {1\over \omega}).
   \label{eq:hnews}
\end{equation}
with $M^{\alpha}=(0,0,M^A)$. An explicit calculation then leads to
\begin{equation}
    N={1\over 4}e^{-2i \delta}\omega^{-2}e^{-2H}F^A F^B
       \{(\partial_u+{\pounds_L})c_{AB}+{1\over 2}c_{AB} D_C L^C
        +2\omega D_A[\omega^{-2}D_B(\omega e^{2H})]\},
     \label{eq:news}
\end{equation}
where $\pounds_L$ denotes the Lie derivative with respect to $L^A$.
The eth versions of these expressions are given in Appendix
\ref{app:appb}.

\subsection{Inertial coordinates} \label{sec:ic}

Equation (\ref{eq:news}) is an expression for the news $N(u,x^A)$ in
terms of the angular coordinates $x^A$ determined by the gauge
conditions on the inner worldtube $\Gamma$. The angular coordinates
$y^A$ of inertial observers at ${\cal I^+}$ are constant along the
generators of ${\cal I^+}$, so they are related to the $x^A$
coordinates by $\tilde n^{\alpha} \partial_{\alpha} y^A =0$ or
\begin{equation}
     (\partial_u + L^B \partial_B) y^A =0. \label{eq:yB}
\end{equation}
We solve this equation for $y^A(u,x^B)$, with the initial condition
$y^A(u_0,x^B)=x^A$, so that $y^A$ are stereographic angular coordinates.
This yields the waveform $N^{\prime}(u,y^A)=N(u,x^B(u,y^A))$ measured
by inertial observers in terms of the retarded time coordinate $u$ in
which the evolution is carried out.

In order to complete the transformation to inertial coordinates
we reexpress $u$ in terms of Bondi time $u_B$, which is an affine
parameter for the generators of ${\cal I}^+$ in a Bondi frame satisfying
\begin{equation}
\tilde n^{\alpha} \partial_{\alpha} u_B =1 \label{eq:uB},
\end{equation}
This allows us to rewrite the news as a complex function 
${\tilde N}(u_B,y^A)=N^{\prime}(u(u_B,y^A),y^A)$
of inertial (Bondi) coordinates on ${\cal I}^+$. 

The news can now be readily decomposed into the two standard
polarization modes. First, we must reexpress the news in terms of a
polarization dyad based upon standard $(\theta,\phi)$ spherical
coordinates rather than upon stereographic coordinates.  The spin
rotation relating the inertial dyads $Q^A_S$ and $Q^A_N$ (respectively belonging to the
south and north stereographic patches of the stereographic
coordinates $y^A$) and the standard dyad ${\tilde Q}^A$
in spherical coordinates 
(${\tilde Q}^A \partial_A =\partial_{\theta}+(i/\sin\theta)\partial_{\phi}$)
is
\begin{eqnarray}
       \tilde Q^A & = & e^{-i \phi} Q^A_S, \\
       \tilde Q^A & = & e^{ i \phi} Q^A_N.
\end{eqnarray}
Since the news function has spin-weight $2$, the transformation
property of spin weighted fields~\cite{eth} leads in each patch to the
news function ${\cal N}$ referred to the the dyad $\tilde Q^A$,
\begin{eqnarray}
 \cal N & = & e^{-2i  \phi} {\tilde N}_S, \\
 \cal N & = & e^{ 2i  \phi} {\tilde N}_N.
\end{eqnarray}

The decomposition of $\cal N$ into real and imaginary parts yields
the standard polarization modes
\begin{eqnarray}
N_{+} & = & \Re[\cal N] \\
N_{\times} & = & \Im[\cal N] .
\end{eqnarray}

\section{Numerical Implementation} \label{sec:num}

In this section we describe a numerical implementation based upon a
second-order accurate finite difference approximation to the equations
presented in Sec \ref{ncf}.

\subsection{The compactified grid}

We introduce a compactified radial coordinate $x=r/(R+r)$, where the
scale factor $R$ is matched to the radius of the world tube. (In the
code runs described here, $R$ is of order 1). The radial coordinate is
discretized as $x_{i}=X+(i-1)\Delta x$ for $i=1\dots N_x$, with $\Delta
x = (1-X) / (N_{x}-1)$. Here setting $X=\frac{1}{2}$ corresponds to
setting $r=R$ at the inner boundary of the grid. The point
$x_{N_{x}}=1$ lies at null infinity. The stereographic coordinate
$\xi=q+ip$ is discretized by $q_{j}=-1 + (j-3)\Delta$ and $p_{k}=-1 + (k-3)\Delta$ for
$j, k = 1 \ldots N_{\xi}$ and $\Delta = 2/(N_{\xi}-5)$. The
evolution proceeds with time step $\Delta u$, subject to a CFL
condition~\cite{cce}.

The fields $J$, $\beta$, $Q$ and $W$ are represented by their values on
this rectangular grid, e.g.  $J^{n}_{ijk}=J(u_{n},x_i,q_{j},p_{k})$.
However, for stability~\cite{cce} the field $U$ is represented by
values at the midpoints $x_{i+\frac{1}{2}} = x_{i}+\Delta x/2$ on a
radially staggered grid (so that
$U^{n}_{ijk}=U(u_{n},x_{i+\frac{1}{2}},q_{j},p_{k})$). In the
following discussion, it is useful to note that asymptotic flatness
implies that the fields $\beta(x)$, $U(x)$, ${\tilde W(x)}= r^{-2}
W(x)$ and $J(x)$ are smooth at future null infinity ${\cal I}^+$ where
$x=1$.

\subsection{Hypersurface equations}

We discretize the hypersurface equations (\ref{eq:beta}) -
(\ref{eq:ww}) along the lines described in \cite{cce}. The only
difference here, is the introduction of the nonlinear terms, which are
evaluated by second order finite differences at the midpoint of the
integration cell.

\subsection{Evolution equation}

The evolution equation (\ref{eq:wev}) has the schematic form
\begin{eqnarray}
    2 \left(rJ\right)_{,ur}
    - \left(r^{-1}V\left(rJ\right)_{,r}\right)_{,r}
    = \frac{J}{r} P_1 + \tilde {\cal H} ,
    \label{eq:wevdis}
\end{eqnarray}
where we have split off the term
\begin{eqnarray}
\tilde {\cal H} &=& - r^{-1}(r^2 \eth U)_{,r}
+ 2 r^{-1} e^{\beta} \eth^2 e^{\beta} -(r^{-1}W)_{,r}J + N_J - \frac{J}{r}P_1
\end{eqnarray}
which contains no $u$-derivatives. In order to obtain an accurate
global discretization, it is convenient to introduce the evolution
variable $\Phi= x J$. ($\Phi$ vanishes in the limit $r\rightarrow 0$
and is smooth at ${\cal I}^+$). With this substitution, the evolution
equation becomes
\begin{eqnarray}
    & & 2 \,( \Phi_{,x} (1 - x) + \Phi)_{,u}
    - {\cal F}_1 ( \Phi_{,x} (1 - x) + \Phi)
     - {\cal F}_2 \Phi_{,xx}  \nonumber \\
    &=& J  (1 - x)  \bigg( \frac{2}{K} \: \Re[ \bar \Phi_{,u}
              (J_{,x} K - J K_{,x} )] - 8\, \beta_{,x}  (1 - x + x \tilde{W}) \bigg)
  + \tilde {\cal H},
    \label{eq:wevdis2}
\end{eqnarray}
where
\begin{eqnarray}
{\cal F}_1 &=&  \tilde W_{,x} x (1 - x) + \tilde W \\
{\cal F}_2 &=& (1 - x)^2 ( (1 - x) + x \tilde W) .
\end{eqnarray}

The elementary computational cell consists of the grid points
$(n,i,k,l)$,  $(n,i\pm 1,k,l)$ and $(n,i-2,k,l)$ on the ``old''
hypersurface and the points $(n+1,i,k,l)$, $(n+1,i-1,k,l)$ and
$(n+1,i-2,k,l)$ on the ``new'' hypersurface. We center the evaluation
at the point $(n+1/2,i-1/2,k,l)$ and approximate the derivatives of
$\Phi$ (to second order). For example, we set
\begin{eqnarray}
\Phi_{,xx} &=& \frac{1}{2 \Delta x^2} ( \Phi^{n+1}_i - 2 \Phi^{n+1}_{i-1}
+\Phi^{n+1}_{i-2} + \Phi^{n}_{i+1} - 2 \Phi^{n}_{i} +\Phi^{n}_{i-1} ), 
\nonumber \\
\Phi_{,x} &=& \frac{1}{2 \Delta x} ( \Phi^{n+1}_i - \Phi^{n+1}_{i-1} +
\Phi^{n}_{i} - \Phi^{n}_{i-1} ) ,  \nonumber \\
\Phi_{,u} &=& \frac{1}{2 \Delta u} ( \Phi^{n+1}_i + \Phi^{n+1}_{i-1} -
\Phi^{n}_{i} - \Phi^{n}_{i-1} ) , \nonumber \\
\Phi_{,xu} &=& \frac{1}{\Delta x \Delta u} \Bigg( (\Phi^{n+1}_i -
\Phi^{n+1}_{i-1}) - \frac{1}{4} (  \Phi^{n}_i - \Phi^{n}_{i-2} +
\Phi^{n}_{i+1} - \Phi^{n}_{i-1} ) \Bigg )  
\label{eq:phixu}
\end{eqnarray}
and
\begin{eqnarray}
\tilde {\cal H} = \frac{1}{2} ( \tilde {\cal H}^{n+1}_{i-1}
 + \tilde {\cal H}^{n}_{i+1} ).
\nonumber
\end{eqnarray}

The evolution to points $(n+1,i,k,l)$ proceeds in an outward march
along the characteristics. However, the variable $\Phi$ appears in the
right hand side of Eq.  (\ref{eq:wevdis2}) so that it is not convenient
to adopt the null parallelogram algorithm used in the quasispherical
case~\cite{cce}.  Instead, we use a Crank-Nicholson \cite{gkobook}
scheme to solve the finite difference version of Eq. (\ref{eq:wevdis2})
for $\Phi^{n+1}_i$.  Rather than implementing an implicit solver to
obtain $\Phi^{n+1}_i$, we use an iterative scheme that ensures second
order accurate evolution.  In addition, the use of 4 points on level
$n$ in the finite differencing of $\Phi_{,xu}$ in Eq. (\ref{eq:phixu})
introduces some numerical dissipation that stabilizes the code even in
the regime of very high amplitude fields.

\subsection{Coordinate transformations}

In order to compute the waveforms $N_{+}$ and
$N_{\times}$measured by inertial observers, we need to carry out the
transformation from the $(u, x^A)$ frame to the $(u_B, y^A)$ frame.  We
achieve this by first transforming to the $(u, y^A)$ and then to $(u_B,
y^A)$, as follows.

\begin{itemize}
\item{Angular transformation:}
Rather than solving the partial differential equation (\ref{eq:yB})
it is more expedient to solve the ordinary differential equation
\begin{equation}
\frac{dx^A}{du} = L^A(u,x^B) \label{eq:transangleode}
\end{equation}
with the initial condition $x^A(u_0) = y^A$ and solutions written as
$x^A(u,y^A)$. This tracks a given curve (the characteristic of the PDE)
along which $y^A = const$ without having to track the entire
congruence.  Since
\begin{eqnarray}
dy^A &=& \partial_u y^A du  + \partial_{x^B} y^A dx^B \nonumber \\
     &=& du ( \partial_u + L^B \partial_{x^B} ) y^A ,
\end{eqnarray}
(\ref{eq:transangleode}) is entirely equivalent to (\ref{eq:yB}).

The numerical integration of (\ref{eq:transangleode}) is
straightforward and is implemented by the centered second order accurate
scheme:
\begin{equation}
(x_A^{n+1} - x_A^{n})_i  = \Delta u \, (L^A)^{(n+\frac{1}{2})}_i .
\end{equation}
Since we are covering the sphere with two stereographic
patches~\cite{eth}, care must be taken to track the $y^A = const$
curves as they pass from one patch to the other. This is accomplished
by the use of a mask which assigns the patch in which the curve lies at
any time step.

\item{Bondi time transformation:}
Similarly, rather than solving the partial differential equation
(\ref{eq:uB}), we determine the transformation to Bondi time by means
of the equivalent ordinary differential equation $du/d u_B={\tilde
n}^{\alpha}\partial_{\alpha} u$, where the differentiation is along the
generators of ${\cal I}^+$. Expressing the left hand side in the
$(u,\ell,x^A)$ coordinate system, this reduces to the differential
equation $du/du_B =\omega^{-1} e^{-2H}$. We set $u_B|_{u=u_0}=u_0$ as the
initial condition.

Again, the integration of this equation is straightforward. We use
the centered second order accurate algorithm
\begin{equation}
(u_B^{n+1} - u_B^{n})_i  = \Delta u \, (\omega e^{2H})^{(n+\frac{1}{2})}_i .
\end{equation}

\item{Inertial news:}
The solution of the above equations determine $x^A(u,y^A)$ and
$u(u_B,y^A)$ and thus allow an explicit transformation of the news from
grid coordinates $(u,x^A)$ to inertial coordinates $(u_B,y^A)$.

\end{itemize}
\section{Numerical Tests} \label{sec:test}

\subsection{Stability}

Numerical experiments were carried out which confirm the {\em
stability} of the code, subject to a CFL condition~\cite{cce}, in the
regime where caustics and horizons do not form. The tests were based
upon long term evolution of highly nonlinear initial null data with
interior worldtube data corresponding to the past horizon of a
Schwarzschild black hole.  See Section~\ref{sec:scat} for a description
of the problem. Stability persists in the regime that the
$u$-direction, along which the grid flows, becomes spacelike. This
occurs routinely in the evolution of ${\cal I}^+$, on which every
tangent direction is spacelike except the null direction of the
generators, $\partial_u + L^B \partial_B$. Furthermore, a test
evolution of the Schwarzschild spacetime in rigidly rotating
coordinates remained stable outside the velocity of light cylinder
where the grid flow becomes superluminal.

\subsection{Accuracy}

We have tested the code for {\em accuracy} and have verified second order
convergence in grid size to analytically known values in the following
test beds which have been established for null codes. 

In all cases, the $L_2$ and $L_{\infty}$ norms of the difference
between the analytical ($F_a$) and numerical ($F_n$) solutions was
calculated for different grid resolutions. These norms were then plotted
against $\Delta x$ and their convergence rate calculated. For
every case, a rate of at least $1.9$ was measured. This procedure is
illustrated in the {\em nonlinear axisymmetric waveforms} case.

We also checked that the relative error $| (F_n - F_a) /F_{a} |$ (when $F_a \neq 0$) behaved as $\kappa \: (\Delta x)^2 $ with $\kappa$ of order 1, thus establishing
the accuracy of the obtained evolution. In particular, for
the case of Schwarzschild in rotating coordinates, $\kappa \leq 0.7 $

\begin{figure}
\centerline{\epsfxsize=6in\epsfbox{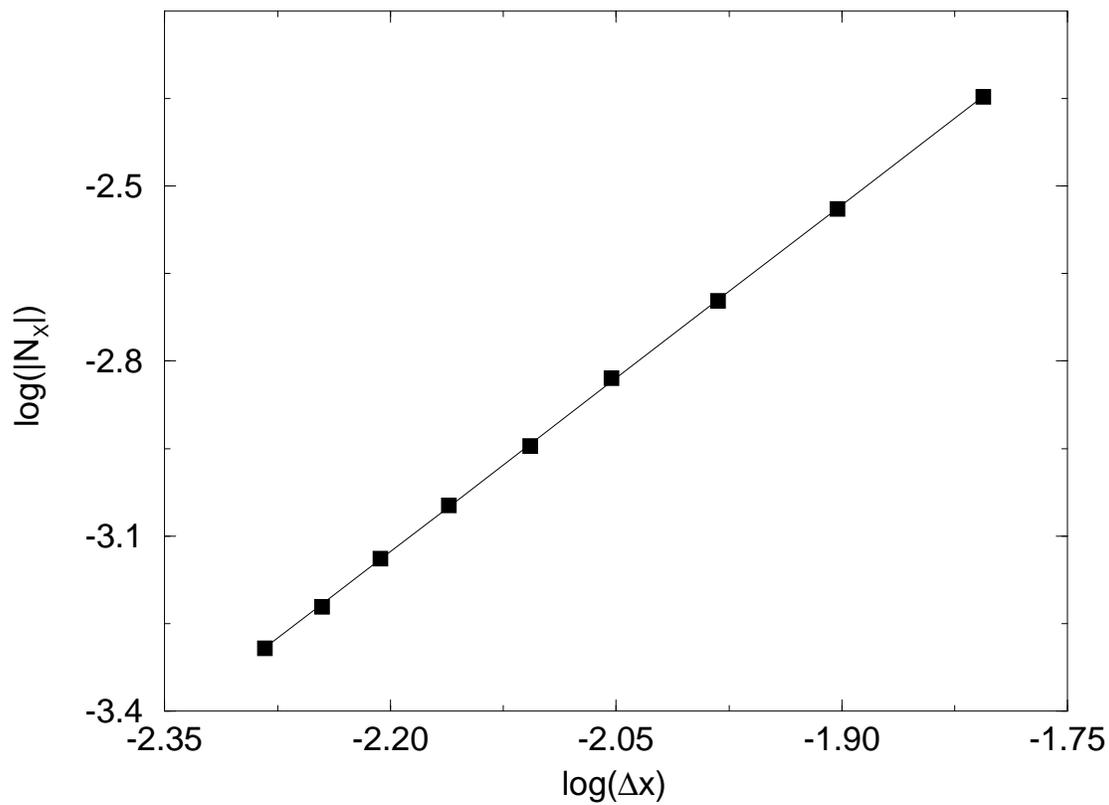}}
\caption{The log of the $L_2$ norm of $N_{\times}$ vs. $log(\Delta x)$. The computation
was performed on grids of size $N_x = 8n + 1$ by $N_{\xi} = 8n + 5$ 
(with $5 \leq n \leq 13$). The corresponding slope is $1.99$, indicating second order convergence.}
\label{fig:conver}
\end{figure}

\begin{itemize}

\item{Linearized Waves:}

We used solutions of the linearized characteristic equations on a
Minkowski background~\cite{cce}. To check accuracy of the code, we
chose a solution representing an outgoing wave with angular momentum
$l=6$ with a very small amplitude ($|J|\simeq10^{-9}$). The inner
boundary was set at $R=1$.  The solution was evolved numerically for
$\Delta u=0.5$ and checked to be second order convergent to the analytic
linearized solution.

\item{Boost and rotation symmetric solutions:}

A family of solutions with exact boost and rotation symmetry called
SIMPLE~\cite{papa} was also used to check accuracy.  Because of their
cylindrical symmetry they are not asymptotically flat but were used to
construct an asymptotically flat solution by smoothly pasting to
asymptotically flat null data outside some radius $R_o$. The resulting
solutions provided a nonlinear test of the second order accuracy of the
evolved field variables in the domain of dependence interior to $R_o$.

\item{Schwarzschild in rotating coordinates:}

By the transformation $\phi \rightarrow \tilde \phi + \omega u$ of the
azimuthal coordinate, the Schwarzschild line element in null
coordinates becomes
\begin{equation}
   ds^2=-(1 - {2 m \over r}-\omega^2 r^2 \sin^2\theta  )du^2 -2 dudr
       +2\omega r^2 \sin^2 \theta dud\phi
       +r^2q_{AB}dx^Adx^B.    \label{eq:smet}
\end{equation}
This gauge transformation gives a nontrivial value for $U$ at ${\cal
I}^+$ and thus  provides a useful test bed to check that the
numerically calculated wave forms remain zero. We evolved this spacetime
from $u=0$ to $u=0.5$ with an inner boundary at $R=3m$ (choosing $m=1$
and $\omega =1$). We confirmed that the news calculated at $u=0.5$
converged to zero to second order. This also checked the accuracy of the
 transformation from code coordinates to inertial coordinates at
${\cal I}^+$.

\item{Waveforms from perturbed Schwarzschild black holes:}

The Robinson-Trautman spacetimes describe a distorted black hole
emitting purely outgoing radiation. The analytic waveform can be easily
calculated in the perturbative regime~\cite{cce} and then compared with
the results from the code. We evolved the spacetime from $u=0$ to
$u=1$, with an inner boundary at $R=3m$ (where the final
black hole has mass $m=1$). Second order convergence of the numerically
obtained news function was confirmed under the $L_{1}$ norm.

\item{Nonlinear axisymmetric waveforms}

The $N_{\times}$ polarization mode must vanish in spacetimes with a
twist-free axisymmetry (the problem originally studied by Bondi).
Since our code uses stereographic coordinates, axisymmetry is not built
in and numerical noise will be produced in the $N_{\times}$ mode. 
We then studied the nonlinear scattering of an $(\ell=2, m=0)$ wave off
a Schwarzschild black hole and determined to what order this noise converged to
zero.
We gave data at $u=0$ and evolved to $u=0.4$. The computation was performed
on grids of size $N_x = 8n + 1$ by $N_{\xi} = 8n + 5$ (with $n \geq 5$).
Figure \ref{fig:conver} shows the plot of $\log(\|N_{\times}\|_{L_2})$ vs. $\log(\Delta x)$. Convergence to second order was verified by the slope of $1.99$

In the next section, we further investigate this scattering
problem in the nonaxisymmetric case.

\end{itemize}

\subsection{Timing}

The characteristic algorithm uses only two complex  and two real
unknowns ($J$ and $U$ and $\beta$ and $W$, respectively). In addition,
the marching structure of the algorithm allows us to carry all
temporary arrays in two dimensional form. As a result, it is much more
computationally efficient than Cauchy algorithms for general
relativity. Of course, being a 3-dimensional algorithm, it cannot be
run in any practical sense on the current generation of workstations or
personal computers.  On one processor of the C90 machine at the
Pittsburgh Supercomputing Center, one time step takes 6 seconds for a
spatial grid of $64^3$ points. A physically complete run, such as the
ones discussed in the next Section, takes 1 hour (running on 1
processor).

\section{Nonlinear scattering off a Schwarzschild black hole}
\label{sec:scat}

The characteristic initial value problem on an outgoing null
hypersurface requires an inner boundary condition on a worldtube (or,
in a limiting case, on the worldline traced out by the vertex of a null
cone).  Some of the worldtube boundary data correspond to gauge
freedoms. As discussed in Section \ref{sec:cons}, when  the worldtube is
timelike this is determined by the lapse and shift arising from a $2+1$
decomposition.  Here we consider an example in which the inner boundary
$\Gamma$ consists of a ingoing nullcone. We adopt the analogue of
setting the shift to zero by choosing coordinates $x^A$ which follow
the ingoing null geodesics. We foliate $\Gamma$ by slices separated by
constant affine parameter $\lambda$ as the analogue of fixing the
lapse. We then complete the specification of the inner boundary by
choosing geometric data on $\Gamma$ to correspond to the ingoing $r=2m$
surface (the past horizon) in a Schwarzschild spacetime. This
determines inner boundary values satisfying the conservation
conditions.

This construction also determines outgoing null coordinates: $u$
which labels the outgoing null hypersurfaces emanating from the the
foliation of $\Gamma$ with $u|_{\Gamma}=\lambda$,  $x^A$ which
assigns stereographic coordinates to the outgoing null rays and the
surface area coordinate $r$.

In these coordinates, the Schwarzschild line element takes the
Eddington-Finklestein form
\begin{eqnarray}
   ds^2=-\left( 1 - {2m \over r} \right)du^2
        -2 dudr
        +r^2q_{AB}dx^Adx^B,
\end{eqnarray}
The initial data corresponds to setting $J=0$ as null data
on $u=0$, with the boundary conditions $\beta = U^A = 0$ and $W = -2 m$
on $\Gamma$ (given by $r=2m$).

We pose the nonlinear problem of gravitational wave scattering off a
Schwarzschild black hole by retaining these boundary conditions on
$\Gamma$ (thus continuing to satisfy the conservation conditions) but choosing null data at $u=0$ corresponding to an incoming
pulse with compact support,
\begin{equation}
   J(u=0,r,x^A) = \left\{ \begin{array}{ll}
  \displaystyle{\lambda \left(1 - \frac{R_a}{r} \right)^4 \,
                        \left(1 - \frac{R_b}{r} \right)^4 \;
                        \sqrt{\frac{4 \pi}{2 l + 1}} \; {}_{2} Y_{l,m}}
                    & \mbox{if $r  \in [R_a,R_b]$} \\
 		    & \\
                  0 & \mbox{otherwise,}
                        \end{array}
                        \right.
\end{equation}
where ${}_{2}Y_{l,m}$ is the spin two spherical harmonic.

\begin{figure}
\centerline{\epsfxsize=6in\epsfbox{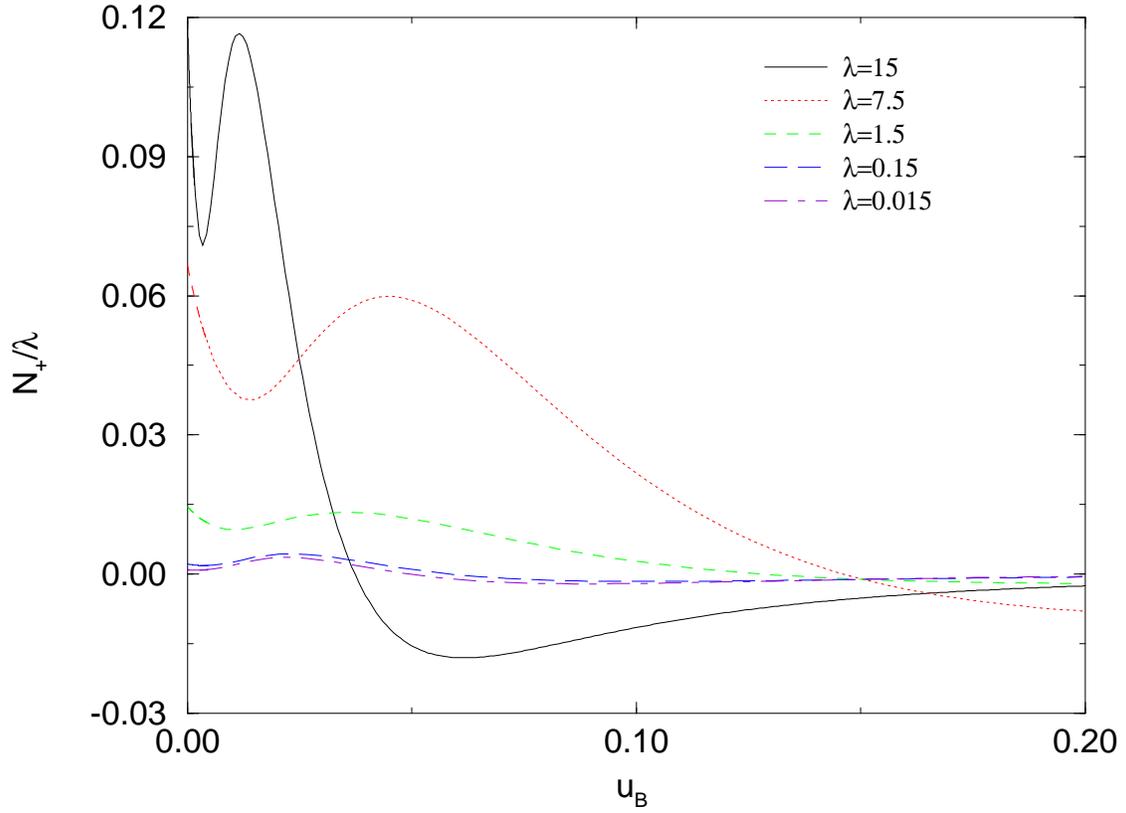}}
\caption{Rescaled $(+)$ component of the news for
$\lambda= 1.5 \times 10^{n}$, with $n=-2,-1,0,1$, and $\lambda= 7.5$ in
a ${}_2Y_{22}$ mode at $\theta=\phi=45^0$. For the smallest values of
$\lambda$, $N_+$ scales linearly with $\lambda$, while at higher values it
shows significant nonlinearity which generates additional modes at early times.}
\label{fig:ampdep}
\end{figure}

We have calculated the plus and cross components of the news function
from $u=0$ to $u=0.2$ (fixing $l=m=2$) for various values of
$\lambda$.  As expected, for small $\lambda$, the news function scales
linearly. However, for larger values, the behavior of the news function shows a
stronger than linear dependence on $\lambda$. This is illustrated in
Figure \ref{fig:ampdep}, which shows the plus component of the news,
rescaled as $N_{+}/\lambda$, versus Bondi time. For $\lambda <
10^{-2}$, the rescaled news are essentially equal. In this perturbative
regime, the news results from the backscattering of the incoming pulse
off the effective potential of the interior Schwarzschild black
hole~\cite{price}. However, for higher values of $\lambda$ the waveform behaves quite differently. Not only is its amplitude greater (up to $3$ times
for $\lambda = 15$) but it also reveals the nonlinear generation of
additional modes at early times for the high amplitude cases. In this
regime, the mass of the system is dominated by the the incoming pulse,
which essentially backscatters off itself in a nonlinear way.

\begin{figure}
\centerline{\epsfxsize=6in\epsfbox{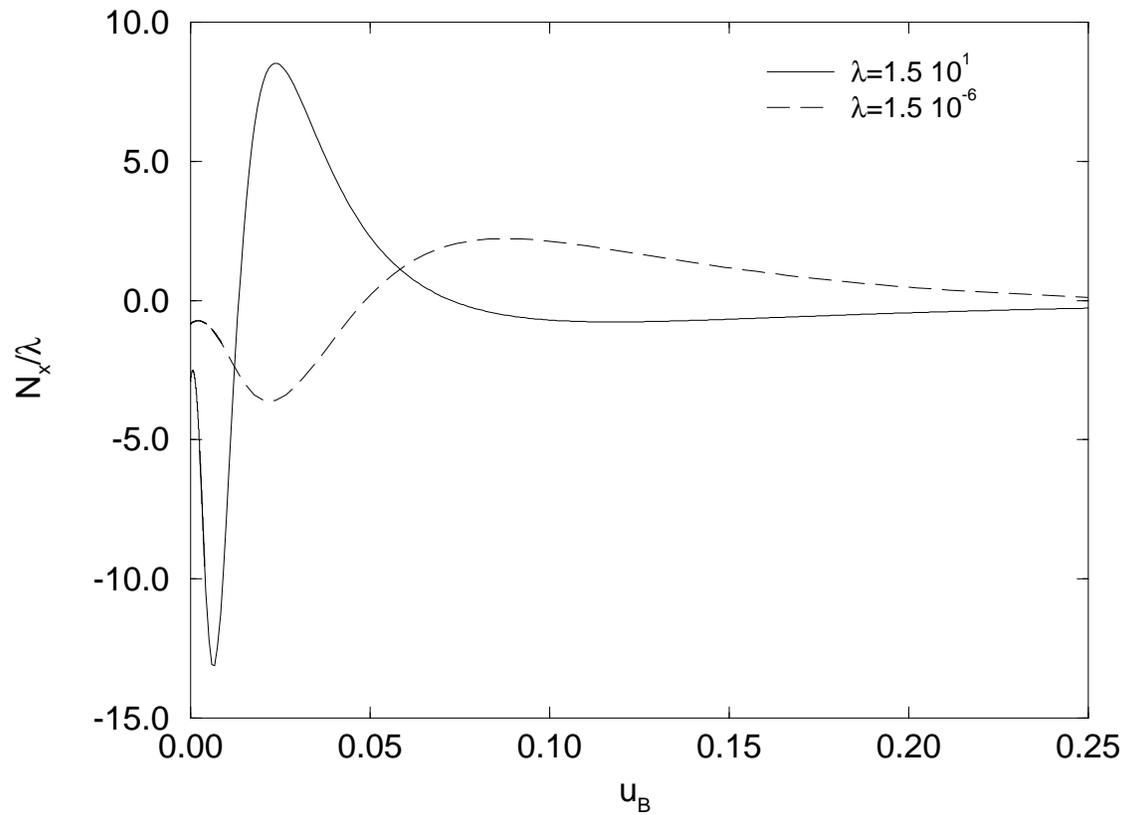}}
\caption{Rescaled $(\times)$ component of the news, for $\lambda= 1.5
\times 10^{n}$, with $n=-6$ and $n=1$ in a ${}_2Y_{22}$ mode at
$\theta=\phi=45^0$.  $N_{\times}/\lambda$ for $\lambda= 15$ differs by
about $300\%$ from the rescaled value for $\lambda=1.5 \times
10^{-6}$.}
\label{fig:newscross}
\end{figure}

The nonlinearity of the news function can be also seen in Figure
\ref{fig:newscross}, which displays $N_{\times}/\lambda$ for amplitudes
$\lambda=1.5 \times 10^{-6}$ and $\lambda=15$. If the scaling were linear,
$| N_{\times}/\lambda |$  would obtain a maximum value of $\approx 4$,
while in fact its maximum is $\approx 13$.

\begin{figure}
\centerline{\epsfxsize=6in\epsfbox{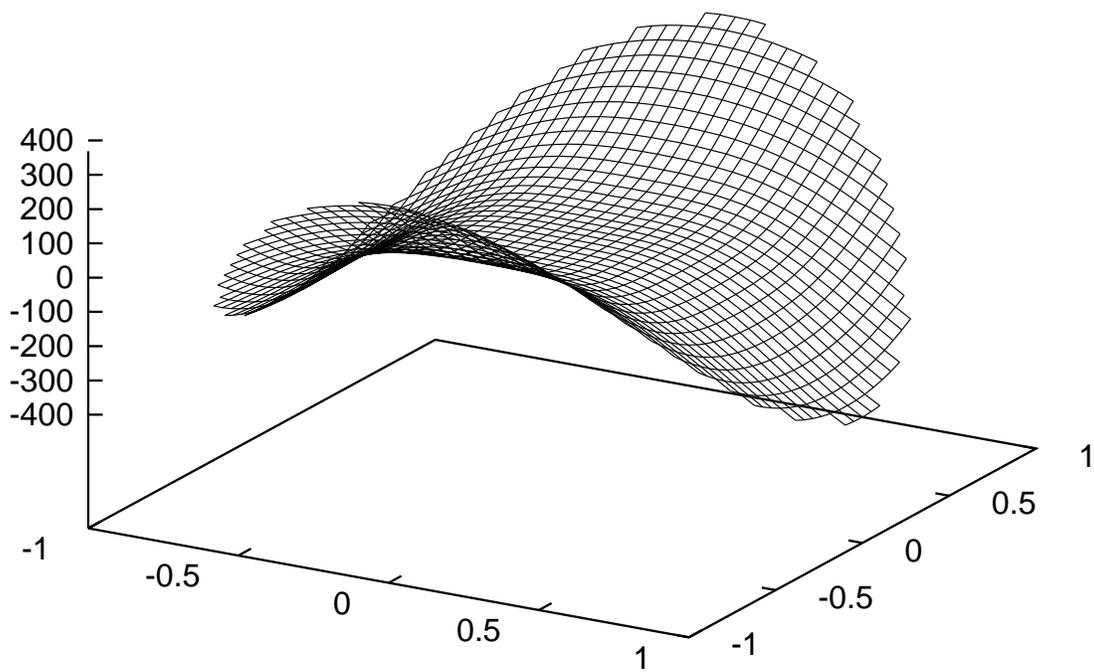}}
\caption{Surface plot showing $N_+(u_B=0.01,y^A)$ on the southern
hemisphere.  The obtained news function remains smooth after performing
the sequence of transformations described in Sec.~\ref{sec:ic} which
takes us from a $(u,x^{A})$ frame to a Bondi frame $(u_{B},y^{A})$.}
\label{fig:surplus}
\end{figure}

It is worth emphasizing that the plus and cross components of the news
function have been obtained by evolving the system in the $(u,x^A)$
coordinate system and then transforming to the $(u_B,y^A)$ coordinates
and to a standard Bondi conformal frame. The surface plot of the plus
component of the news in the southern hemisphere at $u_B= 0.01$, given
in Figure \ref{fig:surplus}, illustrate the smoothness with which our
algorithm accomplishes this complete transformation.

\begin{center}
{\bf ACKNOWLEDGMENTS}
\end{center}

\vspace{0.3cm}
We are grateful to the referee for comments that have improved the
clarity of presentation. This work has been supported by the Binary
Black Hole Grand Challenge Alliance, NSF PHY/ASC 9318152 (ARPA
supplemented) and by NSF PHY 9510895 and NSF INT 9515257 to the
University of Pittsburgh.  N.T.B.  and M.M.  thank the Foundation for
Research Development, South Africa, for financial support, and the
University of Pittsburgh for hospitality. J.W. thanks the Universities
of South Africa and of Durban-Westville for their hospitality. Computer
time for this project has been provided by the Pittsburgh
Supercomputing Center under grant PHY860023P to J.W. and by a grant
from the Center for High Performance Computing of the University of
Texas at Austin, to Richard Matzner.

\appendix

\section{Nonlinear terms} \label{app:appa}

We have the following spin-weighted expressions
for the nonlinear terms in the Einstein equations:

$R_{rr}=0$ leads to Eq. (\ref{eq:beta}) for $\beta_{,r}$ with
\begin{equation}
N_\beta=\frac{r}{8}\left(J_{,r}\bar J_{,r}-K^2_{,r} \right).
\label{eq:nbeta}
\end{equation}

$R_{rA}q^A=0$ leads to Eq's. (\ref{eq:wua}) and (\ref{eq:wq}) for
$U_{,r}$ and $Q_{,r}$with
\begin{equation}
N_U=\frac{e^{2\beta}}{r^2} \left(KQ-Q-J\bar Q \right),
\label{eq:nu}
\end{equation}
and
\begin{eqnarray}
N_Q &=& r^2 \Bigg( (1-K) ( \eth K_{,r} + \bar \eth J_{,r} ) + \eth (\bar J
J_{,r} ) + \bar \eth ( J K_{,r} )  - J_{,r} \bar \eth K \nonumber \\
   & + & \frac{1}{2K^2}(\eth \bar J (J_{,r} - J^2 \bar J_{,r} ) + \eth J (\bar
J_{,r} -\bar J^2  J_{,r} ) ) \Bigg).
\label{eq:nq}
\end{eqnarray}

$R_{AB}h^{AB}=0$ leads to Eq. (\ref{eq:ww}) for $W_{,r}$ with
\begin{eqnarray}
N_W&=& e^{2 \beta} \Bigg( (1-K) ( \eth \bar \eth \beta +  \eth \beta \bar \eth
\beta) + \frac{1}{2} \bigg ( J (\bar \eth \beta)^2 + \bar J (\eth \beta)^2
\bigg) \nonumber \\
& & - \frac{1}{2} \bigg ( \eth \beta ( \bar \eth K - \eth \bar J) + \bar \eth
\beta ( \eth K - \bar \eth J ) \bigg) + \frac{1}{2} \bigg ( J \bar \eth^2 \beta
+ \bar J \eth^2 \beta  \bigg) \Bigg )  \nonumber \\
& & - e^{-2 \beta} \frac{r^4}{8} ( 2 K U_{,r} \bar U_{,r} + J \bar U^2_{,r} +
\bar J U^2_{,r}).
\label{eq:mw}
\end{eqnarray}

$R_{AB}q^Aq^B$ leads to Eq. (\ref{eq:wev}) for $J_{,ur}$ with
\begin{equation}
N_J=N_{J1}+N_{J2}+N_{J3}+N_{J4}+N_{J5}+N_{J6}+N_{J7}+\frac{J}{r}(P_1+P_2
+P_3+P_4)
\label{eq:nj}
\end{equation}
where
\begin{eqnarray}
N_{J1}&=& - \frac{e^{2 \beta}}{r} \bigg ( K ( \eth J \bar \eth \beta + 2 \eth K
\eth \beta - \bar \eth J \eth \beta) + J ( \bar \eth J \bar \eth \beta - 2 \eth
K \bar \eth \beta) - \bar J \eth J \eth \beta \bigg) ,
\nonumber \\
N_{J2}&=& -\frac{1}{2} \bigg ( \eth J ( r \bar U_{,r} + 2 \bar U) + \bar \eth J
( r U_{,r} + 2 U) \bigg) ,
\nonumber \\
N_{J3}&=& (1-K) ( r  \eth U_{,r} + 2 \eth U) - J ( r \eth \bar U_{,r} + 2 \eth
\bar U) ,
\nonumber \\
N_{J4}&=& \frac{r^3}{2} e^{-2 \beta} \bigg( K^2 U^2_{,r} + 2 J K U_{,r} \bar
U_{,r} + J^2 \bar U^2_{,r} \bigg) ,
\nonumber \\
N_{J5}&=& - \frac{r}{2} J_{,r} ( \eth \bar U + \bar \eth U) ,
\nonumber \\
N_{J6}&=& r \Bigg( \frac{1}{2} ( \bar U \eth J + U \bar \eth J ) (J \bar J_{,r}
- \bar J J_{,r} )  \nonumber \\
    & & + ( J K_{,r} - K J_{,r} ) \bar U \bar \eth J
     - \bar U ( \eth J_{,r} - 2 K \eth K J_{,r} + 2 J \eth K K_{,r} ) \nonumber
\\
    & & - U ( \bar \eth J_{,r} - K \eth \bar J J_{,r} + J \eth \bar J  K_{,r} )
\Bigg) ,
\nonumber \\
N_{J7}&=& r ( J_{,r} K -  J K_{,r} ) \bigg ( \bar U ( \bar \eth J - \eth K ) +
U ( \bar \eth K - \eth \bar J ) \nonumber \\
       & & + K ( \bar \eth U - \eth \bar U ) + ( J \bar \eth \bar U - \bar J
\eth U ) \bigg) ,
\nonumber \\
P_1 &=& r^2 \bigg ( \frac{J_{,u}}{K} (\bar J_{,r} K - \bar J K_{,r} ) +
\frac{\bar J_{,u}}{K} ( J_{,r} K -  J K_{,r} ) \bigg)
	    - 8 V \beta_{,r} ,
\nonumber \\
P_2 &=& e^{2 \beta} \Bigg( - 2 K ( \eth \bar \eth \beta + \bar \eth \beta \eth
\beta) - ( \bar \eth \beta \eth K
                     + \eth \beta  \bar \eth K) \nonumber \\
                & + &   \bigg( J ( \bar \eth^2 \beta + (\bar \eth \beta)^2 ) +
                           \bar J ( \eth^2 \beta + (\eth \beta)^2 ) \bigg)
               + ( \bar \eth J \bar \eth \beta + \eth \bar J \eth \beta) \Bigg),
\nonumber \\
P_3 &=& \frac{r}{2} \bigg( ( r \bar \eth U_{,r} + 2 \bar \eth U) +
		       ( r  \eth \bar U_{,r} + 2 \eth \bar U) \bigg) ,
\nonumber \\
P_4 &=& - \frac{r^4}{4} e^{-2 \beta} ( 2 K U_{,r} \bar U_{,r} + J \bar U^2_{,r}
		      + \bar J U^2_{,r} ).
\label{eq:nji}
\end{eqnarray}

\section{Calculating the news in the eth formalism} \label{app:appb}

The eth versions of the expressions needed to calculate the news via
Eq. (\ref{eq:news}) are given below:

\begin{eqnarray}
H^{AB}D_A D_B \log \omega
& = &\frac{1}{4}(
  - 2 \eth^2 \log \omega \bar J
  - 2 \bar \eth^2 \log \omega J
  + 4 \bar \eth \eth \log \omega K
\nonumber \\
& - & \eth \log \omega \eth J \bar J^2
  -  \eth \log \omega \eth \bar J J \bar J
  - 2 \eth \log \omega \eth \bar J
\nonumber \\
& + & 2 \eth \log \omega \eth K \bar J K
  +   \eth \log \omega \bar \eth J \bar J K
  + \eth \log \omega \bar \eth \bar J J K
\nonumber \\
& - & 2 \eth \log \omega \bar \eth K J \bar J
  + \eth J \bar \eth \log \omega \bar J K
  + \eth \bar J \bar \eth \log \omega J K
\nonumber \\
& - & 2 \eth K \bar \eth \log \omega J \bar J
  - \bar \eth \log \omega \bar \eth J J \bar J
  - 2 \bar \eth \log \omega \bar \eth J
\nonumber \\
& - & \bar \eth \log \omega \bar \eth \bar J J^2
  + 2 \bar \eth \log \omega \bar \eth K J K) ,
\end{eqnarray}
\begin{equation}
D_A U^A=(\eth \bar U + \bar \eth U)/2 ,
\end{equation}
\begin{equation}
H_{AC} \bar F^C \partial_u F^A
=( - J_{,u} \bar J K - J_{,u} \bar J + K_{,u} J \bar J
+ K_{,u} K + K_{,u})/(J \bar J + 2 K + 2) .
\end{equation}
For any quantity $\Phi$ having spin-weight zero,
\begin{equation}
     U^A \partial_A \Phi= \frac{1}{2}(\bar U\eth \Phi + U \bar \eth \Phi ) .
\end{equation}
The news $N$ is:
\begin{equation}
N=\frac{1}{4} e^{-2 i \delta} \omega^{-1} A^{-1}
(s_1 + s_2 +\frac{1}{4}(\eth \bar U +\bar \eth U) s_3 -4 \omega^{-2} s_4
+2 \omega^{-1} s_5 ) ,
\end{equation}
where $A= \omega e^{2 \beta}$ and the terms $s_1$ to $s_5$ are:
\begin{eqnarray}
s_1& = &(J^2 \bar J_{,\ell u} + J \bar J J_{,\ell u} - 2 J K K_{,\ell u}
- 2 J K_{,\ell u} + 2 J_{,\ell u} K + 2 J_{,\ell u})/(K + 1) ,
\nonumber \\
\nonumber \\
s_2& =& (\eth J_{,\ell} J \bar J \bar U + 2 \eth J_{,\ell} K \bar U
+ 2 \eth J_{,\ell} \bar U + \eth \bar J_{,\ell} J^2 \bar U \nonumber \\
& - & 2 \eth K_{,\ell} J K \bar U - 2 \eth K_{,\ell} J \bar U
+ 2 \eth U J \bar J K_{,\ell} - 2 \eth U J \bar J_{,\ell} K \nonumber \\
& - & 2 \eth U J \bar J_{,\ell} + 4 \eth U K K_{,\ell} + 4 \eth U K_{,\ell}
+ 2 \eth \bar U J \bar J J_{,\ell} - 2 \eth \bar U J K K_{,\ell} \nonumber \\
& - & 2 \eth \bar U J K_{,\ell} + 4 \eth \bar U J_{,\ell} K
+ 4 \eth \bar U J_{,\ell} + \bar \eth J_{,\ell} J \bar J U
+ 2 \bar \eth J_{,\ell} K U + 2 \bar \eth J_{,\ell} U  \nonumber \\
& + & \bar \eth \bar J_{,\ell} J^2 U - 2 \bar \eth K_{,\ell} J K U
- 2 \bar \eth K_{,\ell} J U + 2 \bar \eth U J^2 \bar J_{,\ell}
- 2 \bar \eth U J K K_{,\ell} \nonumber \\
& - & 2 \bar \eth U J K_{,\ell}
+ 2 \bar \eth \bar U J^2 K_{,\ell} - 2 \bar \eth \bar U J J_{,\ell} K
- 2 \bar \eth \bar U J J_{,\ell})/(2 (K + 1)) ,
\nonumber \\
\nonumber \\
s_3& =& (J^2 \bar J_{,\ell} + J \bar J J_{,\ell} - 2 J K K_{,\ell}
- 2 J K_{,\ell} + 2 J_{,\ell} K + 2 J_{,\ell})/(K + 1) ,
\nonumber \\
\nonumber \\
s_4& =& (\eth A \eth \omega J \bar J + 2 \eth A \eth \omega K
+ 2 \eth A \eth \omega - \eth A \bar \eth \omega J K \nonumber \\
& - & \eth A \bar \eth \omega J - \eth \omega \bar \eth A J K
- \eth \omega \bar \eth A J + \bar \eth A \bar \eth \omega J^2)/(2 (K + 1)) ,
\nonumber \\
\nonumber \\
s_5& =& (2 \eth^2 A J \bar J + 4 \eth^2 A K + 4 \eth^2 A + 2 \bar \eth^2 A J^2
- 4 \bar \eth \eth A J K \nonumber \\
& - & 4 \bar \eth \eth A J + \eth A \eth J J \bar J^2
+ 2 \eth A \eth J \bar J K + 2 \eth A \eth J \bar J
+ \eth A \eth \bar J J^2 \bar J \nonumber \\
& + & 2 \eth A \eth \bar J J K
+ 2 \eth A \eth \bar J J - 2 \eth A \eth K J \bar J K
- 4 \eth A \eth K J \bar J - 4 \eth A \eth K K \nonumber \\
& - & 4 \eth A \eth K
- \eth A \bar \eth J J \bar J K + 2 \eth A \bar \eth J K
+ 2 \eth A \bar \eth J - \eth A \bar \eth \bar J J^2 K \nonumber \\
& + & 2 \eth A \bar \eth K J^2 \bar J - \eth J \bar \eth A J \bar J K
- 2 \eth J \bar \eth A J \bar J - 2 \eth J \bar \eth A K \nonumber \\
& - & 2 \eth J \bar \eth A - \eth \bar J \bar \eth A J^2 K
- 2 \eth \bar J \bar \eth A J^2 + 2 \eth K \bar \eth A J^2 \bar J \nonumber \\
& + & 4 \eth K \bar \eth A J K + 4 \eth K \bar \eth A J
+ \bar \eth A \bar \eth J J^2 \bar J \nonumber \\
& + & \bar \eth A \bar \eth \bar J J^3
- 2 \bar \eth A \bar \eth K J^2 K)/(4 (K + 1)) .
\end{eqnarray}

\end{document}